\documentclass{article}

\usepackage[preprint]{neurips_2024}

\usepackage{footnote}
\usepackage{adjustbox}
\usepackage{wrapfig}
\usepackage[title]{appendix}
\usepackage{wrapfig}
\usepackage{tablefootnote}
\usepackage[utf8]{inputenc} % allow utf-8 input
\usepackage[T1]{fontenc}    % use 8-bit T1 fonts
\usepackage{hyperref}       % hyperlinks
\usepackage{url}            % simple URL typesetting
\usepackage{booktabs}       % professional-quality tables
\usepackage{amsfonts}       % blackboard math symbols
\usepackage{nicefrac}       % compact symbols for 1/2, etc.
\usepackage{microtype}      % microtypography
\usepackage{xcolor}         % colors
\usepackage{graphicx}
\usepackage{subfigure}
\usepackage{amsmath, bm}
\usepackage{makecell}
\usepackage{multirow}
\usepackage{wrapfig}
\usepackage[title]{appendix}
\usepackage{wrapfig}
\usepackage{fancyvrb}
\usepackage{graphicx}

\usepackage{adjustbox}
\usepackage{algorithm}
\usepackage{algpseudocode}

\usepackage{svg}
\usepackage{arydshln}
\usepackage{subcaption}

\makeatletter

\newcommand{\norm}[1]{\left\lVert#1\right\rVert}

\VerbatimFootnotes

\newcommand{\makeappendixtitle}{
  \par
  \begingroup
    \thispagestyle{empty}
    \@makeappendixtitle
  \endgroup
  \let\makeappendixtitle\relax
}

\providecommand{\@makeappendixtitle}{}
\renewcommand{\@makeappendixtitle}{
  \vbox{
    \hsize\textwidth
    \linewidth\hsize
    \vskip 0.1in
    \@toptitlebar
    \centering
    {\LARGE\bf \@title\par}
    \@bottomtitlebar
    \def\And{
      \end{tabular}\hfil\linebreak[0]\hfil%
      \begin{tabular}[t]{c}\bf\rule{\z@}{24\p@}\ignorespaces%
    }
    \def\AND{
      \end{tabular}\hfil\linebreak[4]\hfil%
      \begin{tabular}[t]{c}\bf\rule{\z@}{24\p@}\ignorespaces%
    }
  }
}

\title{DMOSpeech 2: Reinforcement Learning for Duration Prediction in Metric-Optimized Speech Synthesis}

\author{%
  Yingahao Aaron Li$^{1}$\thanks{These authors contributed equally. Correspondence: Y.\,A.~Li (yl4579@columbia.edu).},\enspace
  Xilin Jiang$^{1}$\footnotemark[1],\enspace 
  Fei Tao$^{2}$,\enspace 
  Cheng Niu$^{2}$,\enspace\\[0.25em]
  \textbf{Kaifeng Xu$^{2}$,\enspace Juntong Song$^{2}$,\enspace Nima Mesgarani$^{1}$}\\[0.35em]
  $^{1}$Columbia University,\; $^{2}$NewsBreak\\[0.25em]
}

\begin{document}

\maketitle

\begin{abstract}
  Diffusion-based text-to-speech (TTS) systems have made remarkable progress in zero-shot speech synthesis, yet optimizing all components for perceptual metrics remains challenging. Prior work with DMOSpeech demonstrated direct metric optimization for speech generation components, but duration prediction remained unoptimized. This paper presents DMOSpeech 2, which extends metric optimization to the duration predictor through a reinforcement learning approach. The proposed system implements a novel duration policy framework using group relative preference optimization (GRPO) with speaker similarity and word error rate as reward signals. By optimizing this previously unoptimized component, DMOSpeech 2 creates a more complete metric-optimized synthesis pipeline. Additionally, this paper introduces teacher-guided sampling, a hybrid approach leveraging a teacher model for initial denoising steps before transitioning to the student model, significantly improving output diversity while maintaining efficiency. Comprehensive evaluations demonstrate superior performance across all metrics compared to previous systems, while reducing sampling steps by half without quality degradation. These advances represent a significant step toward speech synthesis systems with metric optimization across multiple components. The audio samples, code and pre-trained models are available at \url{https://dmospeech2.github.io/}. 
\end{abstract}

\section{Introduction}
Text-to-speech (TTS) synthesis has progressed dramatically in recent years, with state-of-the-art systems producing speech virtually indistinguishable from human recordings \citep{tan2024naturalspeech, li2024styletts2, ju2024naturalspeech}. Among the most significant advancements is zero-shot TTS, which is the ability to synthesize speech in the voice of an unseen speaker, given only a short audio sample without speaker-specific training. This capability has transformative potential across applications ranging from personalized digital assistants to accessibility tools and creative content production. 

Despite impressive quality improvements, zero-shot TTS still faces a fundamental challenge: the lack of true end-to-end optimization for perceptual quality metrics. Current approaches struggle to directly optimize key metrics such as speaker similarity and intelligibility in an end-to-end manner, limiting their performance ceiling, especially for smaller and more efficient models. Reinforcement learning (RL) offers a potential indirect optimization approach \cite{chen2024enhancing, zhang2024speechalign, gao2025emo, tian2025preference, hussain2025koel} but comes with significant limitations. The ceiling of RL-based improvement is essentially best-of-N sampling \cite{ichihara2025evaluation}, making its effectiveness heavily dependent on the original model's output diversity. For smaller, more efficient models with limited output diversity, RL may yield minimal improvements. Additionally, traditional RL for TTS imposes substantial computational overhead, as each training step requires generating complete speech samples—often through hundreds of sampling steps—making large-scale training prohibitively expensive without massive computational resources.

As the field has evolved, researchers have pursued two fundamentally different approaches to generating speech, each with their unique hurdles for direct metric optimization. Autoregressive models \cite{wang2023neural, peng2024voicecraft, chen2024vall, wang2024maskgct, du2024cosyvoice, du2025vall, du2024cosyvoice2, zhu2024autoregressive, wang2025spark, song2024touchtts, ye2025llasa} generate speech step-by-step, similar to how large language models produce text. These systems naturally determine the duration of speech during generation but struggle with direct optimization due to the computational expense of backpropagating through their long generation sequences. While RL could theoretically help, these sequential models only amplify the previously mentioned limitations of RL approaches. Meanwhile, diffusion-based systems \citep{le2024voicebox, shen2023naturalspeech, li2024styletts, eskimez2024e2, yang2024simplespeech, lee2024ditto, chen2024f5} take a different approach, treating speech synthesis as an inpainting task that requires knowing the total speech duration in advance. This creates a natural division in the pipeline: first predicting how long the speech should be, then generating the actual audio content. The challenge here is not just computational but also structural. Without a differentiable connection between these two components, traditional optimization techniques cannot flow through the entire system. Research has demonstrated that input durations significantly impact key metrics like speaker similarity (SIM) and word error rate (WER) \citep{eskimez2024e2}, yet existing systems either train duration predictors separately from speech generation \citep{le2024voicebox, lee2024ditto} or use heuristic approaches based on prompt speaking rates \citep{chen2024f5, eskimez2024e2}.

\begin{figure*}[!t]
\centering
\includegraphics[width=\linewidth]{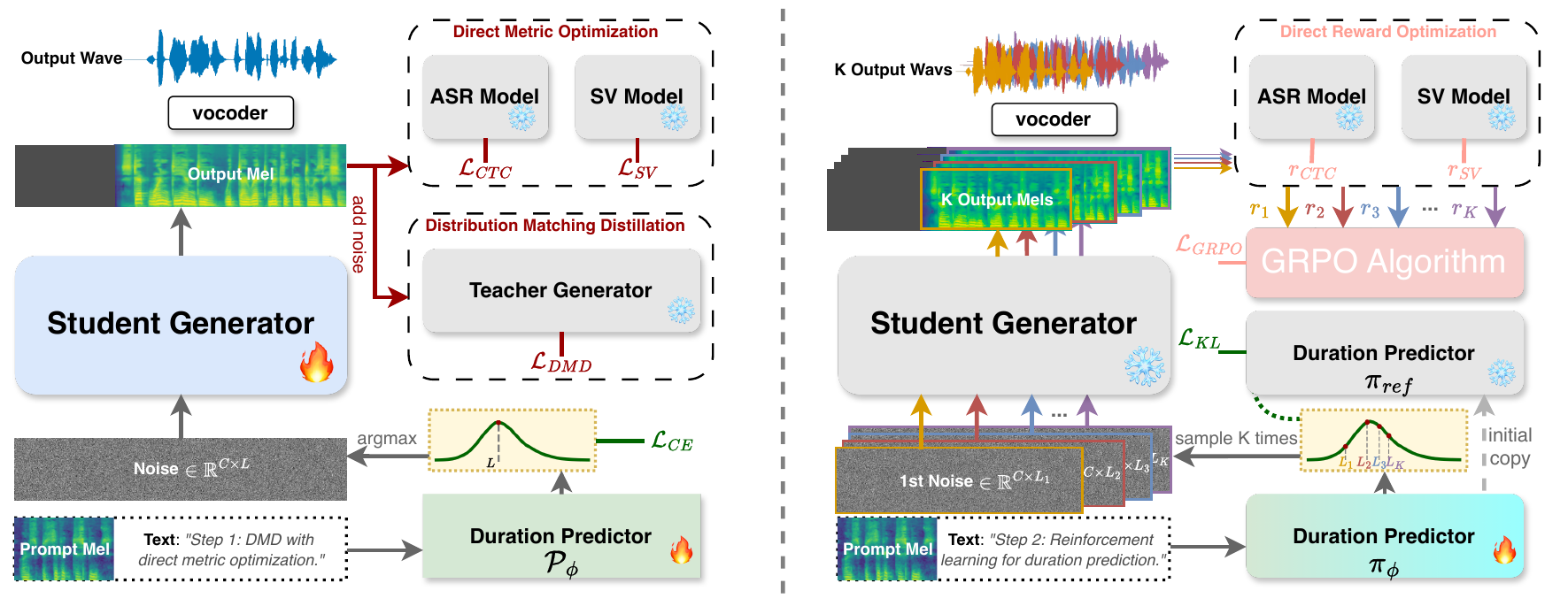}
\caption{Overview of the DMOSpeech 2 framework. \textbf{(a) Left:} The original DMOSpeech architecture, where the duration predictor ($\mathcal{P}_\phi$) is trained self-supervisedly and separate from the TTS component, creating a disconnection that prevents end-to-end optimization. \textbf{(b) Right:} Our proposed DMOSpeech 2 framework, which employs Group Relative Policy Optimization (GRPO) to train the duration predictor with reinforcement learning (Algorithm \ref{alg:GRPO}), using speaker similarity and word error rate as reward signals, enabling end-to-end optimization of the entire TTS pipeline.}
\label{fig:overview}
\vspace{-10pt}
\end{figure*}

The original \textbf{D}irect \textbf{M}etric \textbf{O}ptimization \textbf{Speech} framework  \citep{li2024dmospeech} made significant by enabling direct metric optimization for the speech generation component through diffusion model distillation. By reducing sampling steps from 128 to 4 and establishing direct gradient pathways within the generation process, DMOSpeech enabled direct optimization for speaker similarity and intelligibility. However, a critical limitation remained: the duration predictor component was still outside the optimization loop, creating a bottleneck in overall system quality.

This paper introduces \textbf{DMOSpeech 2}, which addresses the duration prediction challenge through reinforcement learning. We propose modeling the duration predictor as a probabilistic policy and applying reinforcement learning with group relative policy optimization (GRPO), using speaker similarity and word error rate as reward signals. Importantly, by applying RL specifically to the duration predictor and operating on samples generated by our efficient 4-step student model, we dramatically reduce the computational overhead typically associated with RL for TTS. This targeted approach also side-steps the limitations of whole-system RL, as optimizing duration prediction is a much more constrained problem than optimizing speech generation directly.

Additionally, to address the output diversity reduction observed in the original DMOSpeech as a consequence of distribution matching distillation \citep{yin2024one}, we introduce teacher-guided sampling, a hybrid approach that leverages the teacher model for initial denoising steps before transitioning to the student model. This strategy restores diversity to near-teacher levels while still achieving a 2$\times$ reduction in sampling steps and maintaining the significant quality improvements enabled by our direct metric optimization approach.

Using the flow-matching-based F5-TTS \citep{chen2024f5} as our teacher model, our comprehensive evaluations demonstrate that DMOSpeech 2 significantly outperforms both the previous system and other recent baselines across all metrics. The reinforcement learning approach to duration prediction results in particularly notable improvements in speaker similarity and word error rate, precisely targeting the limitations identified in previous systems.

The contributions of this work are twofold: 1) we propose a computationally efficient reinforcement learning framework specifically for duration prediction in non-parallel TTS systems, enabling alignment with perceptual metrics without the overhead typically associated with RL approaches, and 2) we propose a teacher-guided sampling for diffusion model distillation, restoring output diversity while maintaining computational efficiency. We will also make the source code and pre-trained models publicly available for future research in the community.

\section{Related Works}
\textbf{Zero-Shot Text-to-Speech Synthesis} \quad Zero-shot TTS has evolved significantly over recent years, with approaches broadly categorized into two main paradigms. Early methods relied on speaker embeddings from pre-trained encoders \citep{casanova2022yourtts, casanova2021sc, wu2022adaspeech, lee2022hierspeech} or end-to-end speaker encoders \citep{li2024styletts2, min2021meta, li2022styletts, choi2022nansy++}, but struggled with generalization due to their dependence on extensive feature engineering and with direct metric optimization due to their non-differentiable components such as duration predictors. Recent advancements have primarily focused on prompt-based approaches, which can be divided into autoregressive and diffusion-based methods. Autoregressive models \citep{wang2023neural, peng2024voicecraft, chen2024vall, wang2024maskgct, du2024cosyvoice, du2025vall, du2024cosyvoice2, zhu2024autoregressive, wang2025spark, song2024touchtts, ye2025llasa} generate speech sequentially and naturally determine duration during generation, but face limitations in direct optimization due to the computational expense of backpropagation through long generation sequences. In contrast, diffusion-based approaches \citep{le2024voicebox, shen2023naturalspeech, li2024styletts, eskimez2024e2, yang2024simplespeech, lee2024ditto, chen2024f5} treat speech synthesis as an inpainting task requiring predetermined speech duration, creating a natural division between duration prediction and actual speech generation. Although DMOSpeech \citep{li2024dmospeech} made progress by enabling direct optimization for the speech generation component, it still left the duration predictor outside the optimization loop. While duration inputs significantly impact metrics like speaker similarity and word error rate \citep{eskimez2024e2}, existing systems either train duration predictors separately \citep{le2024voicebox, lee2024ditto} or use heuristic approaches based on prompt speaking rates \citep{chen2024f5, eskimez2024e2}. In DMOSpeech 2, we optimize the previously unoptimized duration predictor with reinforcement learning for perceptually relevant metrics. 

\noindent \textbf{Reinforcement Learning in Speech Synthesis} \quad Reinforcement learning (RL) has emerged as a promising approach for aligning speech synthesis systems with human perceptions, though its application to TTS presents unique challenges. Recent work has explored various RL techniques for improving TTS quality. SpeechAlign \citep{zhang2024speechalign} introduced an iterative self-improvement strategy for neural codec language models that constructs preference datasets and optimizes toward human preferences. Similarly, UNO \citep{chen2024enhancing} proposed an uncertainty-aware optimization framework that integrates subjective human evaluation directly into the TTS training loop without requiring a separate reward model. Several approaches have focused on specific aspects of speech quality: \citep{gao2025emo} developed Emo-DPO for controllable emotional speech synthesis, differentiating subtle emotional nuances through preference optimization, while \citep{tian2025preference} demonstrated that direct preference optimization (DPO) consistently improves intelligibility and speaker similarity in LM-based TTS. Koel-TTS \citep{hussain2025koel} enhanced encoder-decoder TTS models through preference alignment guided by automatic speech recognition and speaker verification. For diffusion-based TTS specifically, \citep{chen2024reinforcement} introduced diffusion model loss-guided RL policy optimization (DLPO) to improve naturalness and quality, and \cite{sun2025f5r} employed group relative policy optimization for flow-matching-based TTS models. However, these approaches incur substantial computational overhead, as each training step requires generating complete speech samples, often through hundreds of sampling steps, making large-scale training prohibitively expensive. Additionally, the effectiveness of RL is heavily dependent on the original model's output diversity, potentially yielding minimal improvements for smaller, more efficient models with limited diversity. Most existing approaches apply RL to the entire TTS pipeline, which exacerbates these challenges. DMOSpeech 2 addresses these limitations by specifically targeting RL to the duration predictor component, dramatically reducing computational overhead by operating on samples generated through an efficient 4-step student model, while simultaneously addressing the critical optimization gap in current non-parallel zero-shot TTS systems.

\section{Methods}

\begin{figure*}[!t]
\centering
\includegraphics[width=\textwidth]{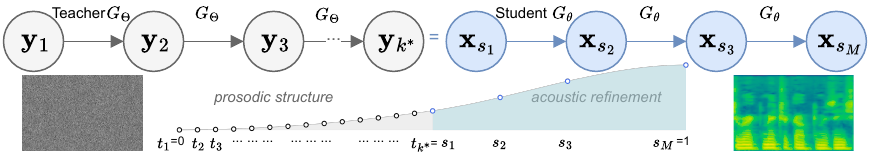}
\caption{Illustration of teacher-guided sampling (Algorithm \ref{alg:teacher_guided_sampling}). The process begins with noise and uses the teacher model $G_{\Theta}$ for early denoising steps (gray circles) to establish prosodic structure up to a transition point $t_{k^*}$. Then, the student model $G_\theta$ (blue circles) takes over for the remaining steps to refine acoustic details in much fewer steps. }
\label{fig:teacher}
\vspace{-10pt}
\end{figure*}

\subsection{DMOSpeech with Flow Matching}

DMOSpeech \citep{li2024dmospeech} is a framework for efficient zero-shot TTS that combines distribution matching distillation \cite{yin2024one} with direct metric optimization. DMOSpeech 2 builds upon the original DMOSpeech framework while adopting F5-TTS \citep{chen2024f5} as the teacher model. This section summarizes the key components of our approach, highlighting the adaptations made for flow matching-based models. Fig.~\ref{fig:overview}a illustrates the DMOSpeech architecture with details in Appendix \ref{app:B}. 

Unlike the original DMOSpeech which operated on latent representations from an audio autoencoder, DMOSpeech 2 directly generates mel-spectrograms, with waveforms synthesized using the pre-trained Vocos \cite{siuzdak2023vocos} vocoder. The framework consists of three training components. First, a student generator $G_\theta$ is trained through improved distribution matching distillation (DMD 2) \citep{yin2024improved} to match a pre-trained teacher model in distribution. This allows the student to generate high-quality speech with significantly fewer sampling steps (4 steps). Second, multi-modal adversarial training with a discriminator improves the perceptual quality of the generated speech. Finally, the direct metric optimization component enables end-to-end optimization of word error rate and speaker similarity metrics with pre-trained automatic speech recognition (ASR) models and speaker verification (SV) models on mel-spectrograms.  

During inference, DMOSpeech generates speech directly from noise in four denoising steps, conditioned on the input text and speaker prompt and the total duration of the target speech. The process begins with sampling Gaussian noise $\mathbf{z} \sim \mathcal{N}(0, I)$ at a predefined duration $L$, which is determined by a separate duration predictor. The student generator $G_\theta$ then transforms this noise into mel-spectrograms through four sequential steps using the sway sampling schedule \cite{chen2024f5} with coefficient $u = -1$ at noise levels $t \in \{0.0000, 0.0761, 0.2929, 0.6173\}$ rather than uniform steps. The final spectrograms are converted to waveforms using the vocoder.

While DMOSpeech enabled direct metric optimization for the generator, it still maintained a critical limitation: the duration predictor remained outside the optimization loop. DMOSpeech 2 addresses this limitation through reinforcement learning, as detailed in the following sections.

\subsection{Speech Length Predictor with RL}

As established in the previous section, while DMOSpeech enables direct optimization of the speech generator, a critical limitation remains: the duration predictor sits outside the optimization loop, creating a disconnection that prevents end-to-end gradient-based optimization. This separation is particularly problematic because speech duration significantly impacts perceptual metrics like speaker similarity (SIM) and word error rate (WER) \citep{eskimez2024e2}. To address this limitation, DMOSpeech 2 introduces a novel reinforcement learning approach specifically targeting the speech length predictor.

\subsubsection{\textbf{Duration Predictor Architecture}}
We adopt an encoder-decoder transformer architecture similar to DiTTo-TTS \citep{lee2024ditto} for our speech length predictor. Unlike conventional duration models that predict phoneme-level durations, our model is specifically designed to predict the total remaining length of speech to be generated.

Formally, let $\mathbf{x}$ represent the input text sequence and $\mathbf{p}_t$ represent the speech prompt up to frame $t$. Our speech length predictor $\mathcal{P}_\phi$ with parameters $\phi$ is trained to predict $L_t$, which is the number of remaining frames needed to complete the utterance:
\begin{equation}
    P_\phi(L_t | \mathbf{x}, \mathbf{p}_t) = \mathcal{P}_\phi(\mathbf{x}, \mathbf{p}_t),
\end{equation}
\noindent where $L_t$ represents the length of the speech segment from frame $t$ to the end. This formulation creates an autoregressive structure where the predicted remaining length decreases as the speech prompt extends. The architecture consists of a bidirectional text encoder that processes the input text to capture comprehensive contextual information. The decoder, equipped with causal masking to prevent future lookahead, takes the mel-spectrogram of the speech prompt as input. Cross-attention mechanisms integrate text features from the encoder, and the final layer applies softmax activation to predict a distribution over possible remaining lengths within a predefined maximum length. Our implementation uses a transformer with 4 encoder layers for text processing and 4 decoder layers with cross-attention mechanisms. The model employs 8 attention heads in each layer with a hidden dimension of 512. We set the maximum total duration to be 30 seconds binned into 300 possible duration classes, with increments of 100 ms. 

During training, the ground truth label for the remaining audio length decreases by one at each subsequent time step. For a batch of sequences with mel-spectrogram lengths $\{L_1, L_2, ..., L_B\}$, where $B$ is the batch size, the target remaining length is a decreasing sequence $(L_i-1, L_i-2, ..., 1, 0)$ for each training example $L_i$. The predictor is initially trained separately from the flow-matching model using cross-entropy loss between the predicted distribution and the ground truth remaining lengths. In DMOSpeech 2, we extend this training process with reinforcement learning to directly optimize for perceptual quality metrics.

\subsubsection{\textbf{GRPO-based Duration Optimization}}

To enable direct optimization for perceptual metrics, we formulate the speech length predictor as a stochastic policy in a reinforcement learning framework and apply group relative policy optimization (GRPO) \citep{shao2024deepseekmath}, which allows us to optimize the length predictor directly for perceptual metrics without need of a differentiable pathway to the generator. The detailed algorithm is provided in Algorithm \ref{alg:GRPO}.

For each training instance $\mathbf{x}$ the input text and $\mathbf{p}$ the prompt, we define the policy for predicting the total speech length $\pi_\phi(L|\mathbf{x},\mathbf{p}) = \mathcal{P}_\phi(\mathbf{x}, \mathbf{p})$. During training, we sample $K$ different duration predictions for each input, where $K$ is the group size:
\begin{equation}
    L_k \sim \pi_\phi(L|\mathbf{x},\mathbf{p}), \quad k = 1, 2, ..., K,
\end{equation}
For each sampled duration, we generate speech using our efficient 4-step student model:
\begin{equation}
    \mathbf{y}_k = G_\theta(\mathbf{z}, \mathbf{x}, \mathbf{p}, L_k), \quad z \sim \mathcal{N}(0, I),
\end{equation}
where $G_\theta$ is our student generator and $\mathbf{z}$ is the initial noise. We then compute rewards for each generated speech sample using a combination of speaker similarity and speech recognition metrics:
\begin{equation}
    r_k = \log p(\mathbf{x} | C(\mathbf{y}_k)) + \lambda_{\text{SIM}} \cdot \frac{\mathbf{e}_{\text{p}} \cdot \mathbf{e}_{y_k}}{\norm{\mathbf{e}_{\text{p}}} \norm{\mathbf{e}_{y_k}}},
\end{equation}
where $C(\cdot)$ is a pre-trained CTC-based ASR model operating on mel-spectrograms, $\mathbf{e}_{\text{p}} = {S}(\mathbf{p})$ and $\mathbf{e}_{y_k} = {S}(\mathbf{y}_k)$ are the speaker embeddings of the prompt and student-generated speech, and $\lambda_{\text{SIM}}$ is the weighting factor. We chose $\lambda_{\text{SIM}} = 3$ to balance the contributions from the embedding similarity and word error rate (see Appendix \ref{app:rl_hyperparams} for detailed discussion).

We normalize the reward to compute the advantage:
\begin{equation}
    A_k = \frac{r_k - \mu_r}{\sigma_r },
\end{equation}
where $\mu_r$ and $\sigma_r$ are the mean and standard deviation of rewards within the group. 

In GRPO, we maintain three distinct policies. The current policy $\pi_\phi$ is the speech length predictor being actively trained. The old policy $\pi_{\text{old}}$ is the version of the policy from which the current batch of samples was generated. In practice, this is typically the policy from several optimization steps ago. The reference policy $\pi_{\text{ref}}$ is a frozen copy of the initially supervised model created at the beginning of RL training and kept constant throughout the process to serve as an anchor for regularization. We define the ratio $R_k$ as :
\begin{equation}
    R_k = \frac{\pi_\phi(L_k|\mathbf{x},\mathbf{p})}{\pi_{\text{old}}(L_k|\mathbf{x},\mathbf{p})}
\end{equation}

\noindent The GRPO loss for a single sample is: 
\begin{equation}
\mathcal{L}_k = \min\left(A_k \cdot R_k, A_k \cdot \text{clip}\left(R_k, 1-\varepsilon, 1+\varepsilon\right) \right) - \beta \cdot \text{KL}
\end{equation}
\noindent where $\varepsilon = 0.2$ is the clipping parameter that limits the policy update magnitude, $\beta = 0.04$ controls the strength of KL regularization, and $\text{KL} = \mathbb{D}_{\text{KL}}[\pi_\phi||\pi_{\text{ref}}]$ is the KL divergence between the current policy and the reference policy, preventing the trained policy from deviating too far from the initial model. 

The full GRPO loss is defined as: 
\begin{equation}
\mathcal{L}_{\text{GRPO}} = -\mathbb{E}_{\mathbf{x}, \mathbf{p}} \left[ \frac{1}{K} \sum\limits_{k=1}^K \mathcal{L}_k \right]
\end{equation}

A challenge in applying RL to duration prediction is the potential for sparse rewards and limited exploration. If the model consistently predicts similar durations, it may fail to discover potentially superior alternatives. To address this, we incorporate temperature-based exploration during sampling. The Gumbel-softmax temperature parameter $\tau$ (set to 0.7 in our implementation) controls the entropy of the length distribution, with higher temperatures encouraging exploration of diverse length predictions:

\begin{equation}
    \pi_\phi^\tau(L|\mathbf{x},\mathbf{p}) = \frac{\exp(\log \pi_\phi(L|\mathbf{x},\mathbf{p}) / \tau)}{\sum_{L'} \exp(\log \pi_\phi(L'|\mathbf{x},\mathbf{p}) / \tau)}
\end{equation}

We also implement a quality control mechanism that skips batches with insufficient reward diversity ($\max(r) - \min(r) < 0.01$), ensuring that the model only learns from batches where meaningful distinctions between good and bad duration predictions can be made. This approach prevents wasting computational resources on batches where all sampled durations yield similar quality speech, focusing training on examples where optimization can make a significant difference.

\begin{algorithm}[!t]
\caption{GRPO-based Speech Length Predictor Training}
\label{alg:GRPO}
\begin{algorithmic}[1]
\State Initialize speech length predictor $\mathcal{P}_\phi$ with supervised training
\State Create reference model $\pi_{\text{ref}}$ as a frozen copy of initial model
\State Initialize batch queue $\mathcal{Q} \leftarrow []$
\For{step = 1 to max\_steps}
    \While{size($\mathcal{Q}$) < 5}
        \State Sample batch $(\mathbf{x}, \mathbf{p})$ from dataset
        \State Compute policy from model: $\pi_\phi \leftarrow \mathcal{P}_\phi(\mathbf{x}, \mathbf{p})$
        \For{$k = 1$ to $K$}
            \State $L_k \sim F_{\text{Gumbel}}(\pi_\phi, \tau)$
            \State $\mathbf{y}_k \leftarrow G_\theta(\mathbf{z}, \mathbf{x}, \mathbf{p}, L_k)$
            \State $r_k \leftarrow \log p(\mathbf{x} | C(\mathbf{y}_k)) + \lambda_{\text{SIM}} \cdot \frac{\mathbf{e}_{\text{p}} \cdot \mathbf{e}_{y_k}}{\norm{\mathbf{e}_{\text{p}}} \norm{\mathbf{e}_{y_k}}}$
        \EndFor
        \If{$\max(r) - \min(r) > 0.01$}
            \For{$k = 1$ to $K$}
                \State $A_k \leftarrow \frac{r_k - \mu_r}{\sigma_r}$
            \EndFor
        \Else
            \State \textbf{continue}
        \EndIf
        \State $\pi_{\text{old}} \leftarrow \pi_\phi$
        \State Push $([A_1, \ldots, A_K], [L_1, \ldots, L_K], \pi_{\text{old}})$ to $\mathcal{Q}$ 
    \EndWhile
    \State Dequeue $([A_1, \ldots, A_K], [L_1, \ldots, L_K], \pi_{\text{old}})$  from $\mathcal{Q}$
    \State $\pi_\phi  \leftarrow \mathcal{P}_\phi(\mathbf{x}, \mathbf{p})$
    \State $\text{KL} \leftarrow \mathbb{D}_{\text{KL}}[\pi_\phi||\pi_{\text{ref}}]$
    \State Initialize loss $\mathcal{L} \leftarrow 0$
    \For{$k = 1$ to $K$}
            \State $R_k \leftarrow \frac{\pi_\phi(L_k|\mathbf{x},\mathbf{p})}{\pi_{\text{old}}(L_k|\mathbf{x},\mathbf{p})}$
            \State $R_{\text{clipped}}  \leftarrow \text{clip}(R_k, 1-\varepsilon, 1+\varepsilon)$
            \State $\mathcal{L} \leftarrow \mathcal{L} - \frac{1}{K}(\min(A_k \cdot R_k, A_k \cdot R_{\text{clipped}}) - \beta \cdot \text{KL})$
    \EndFor  
    \State Update model parameters with gradient of $\mathcal{L}$
\EndFor
\end{algorithmic}
\end{algorithm}

\begin{algorithm}[!t]
\caption{Teacher-Guided Sampling}
\label{alg:teacher_guided_sampling}
\begin{algorithmic}[1]
\Require Teacher model $G_{\Theta}$, student model $G_{\theta}$, teacher steps $K$, student steps $M$, switching time $t_{\text{switch}}$, text embedding $\mathbf{x}$, prompt embedding $\mathbf{p}$, duration $L$, CFG strength $\lambda$
\State Sample $\mathbf{z} \sim \mathcal{N}(0, \mathbf{I})$ with length $L$
\State Initialize $\mathbf{y}_0 \leftarrow \mathbf{z}$
\State Generate teacher time steps $\{t_1, t_2, \ldots, t_K\}$ using sway sampling, $t_1 = 0$
\State Find index $k^*$ such that $t_{k^*} \leq t_{\text{switch}} < t_{k^*+1}$
\For{$k = 1$ to $k^*$}
    \State $\mathbf{v}_{k} \leftarrow G_{\Theta}(\mathbf{y}_{k-1}, \mathbf{x}, \mathbf{p}, t_k)$
    \State $\mathbf{y}_k \leftarrow \mathbf{y}_{k-1} + (t_k - t_{k-1}) \cdot \mathbf{v}_k$
\EndFor
\State Generate student time steps $\{s_1, s_2, \ldots, s_M\}$ with $s_1 = t_{k^*}$ and $s_M = 1$
\State $\mathbf{x}_{s_1} \leftarrow \mathbf{y}_{k^*}$
\For{$m = 1$ to $M$}
    \State $\hat{\mathbf{x}}_1^m \leftarrow G_{\theta}(\mathbf{x}_{s_m}\,; \mathbf{x}, \mathbf{p}, s_m)$
    \If{$m < M$}
        \State Sample $\boldsymbol{\epsilon} \sim \mathcal{N}(0, \mathbf{I})$
        \State $\mathbf{x}_{s_{m+1}} \leftarrow (1-s_{m+1}) \boldsymbol{\epsilon} + s_{m+1} \hat{\mathbf{x}}_1^m$
    \EndIf
\EndFor
\State $\hat{\mathbf{x}}_1^M \leftarrow G_{\theta}(\mathbf{x}_{s_M}\,; \mathbf{x}, \mathbf{p}, s_M)$\\
\Return $\hat{\mathbf{x}}_1^M$
\end{algorithmic}
\end{algorithm}

\subsection{Teacher-Guided Sampling}

\subsubsection{\textbf{Mode Shrinkage in Distribution Matching Distillation}}

One notable limitation of distribution matching distillation observed in the original DMOSpeech is a phenomenon we refer to as \textit{mode shrinkage}. When student models are trained to generate speech in significantly fewer steps than their teacher, they tend to focus on high-probability regions of the data distribution, reducing diversity of the generated samples. While the student model exhibits similar mode coverage in sound quality compared to the teacher as indicated by the UTMOS \cite{saeki2022utmos} distributions, it demonstrates less diversity in prosodic features such as intonation patterns, rhythm variations, and speech cadences (Figure \ref{fig:diversity_comparison}). This suggests that diversity reduction primarily occurs in the temporal and structural dimensions of speech rather than in its spectral characteristics.

The root cause of this diversity reduction can be traced to the diffusion process dynamics. In diffusion-based speech synthesis, different noise levels correspond to distinct aspects of the speech generation process. At high noise levels (early denoising steps), the model primarily establishes prosodic elements, phoneme durations, pauses, pitch contours, and text-speech alignments, essentially the semantic and structural framework of the utterance. In contrast, at low noise levels (later denoising steps), the model refines acoustic details such as voice quality, speaker identity, and spectral characteristics. When the student model is constrained to generate speech in just a few steps, it necessarily compresses this hierarchical generation process. Our empirical observations suggest that this compression disproportionately affects the diversity of prosodic and structural elements established in the early denoising phase.

\subsubsection{\textbf{Hybrid Sampling Strategy}}

To address the mode shrinkage problem, we introduce \textit{teacher-guided sampling}, a hybrid approach that leverages the teacher model's diversity while preserving the student model's efficiency and improved speaker similarity from direct metric optimization. The core insight of our approach is to exploit the natural division of labor in the diffusion process: use the teacher model for early denoising steps on prosodic structure and the student model for acoustic refinement of later steps. Specifically, we employ the teacher model to perform the initial denoising steps up to a predefined noise level $t_{\text{switch}}$, which establihes diverse prosodic patterns and text-speech duration alignments. Then, we switch to the student model, which completes the remaining denoising process from $t_{\text{switch}}$ to 1 in just a few efficient steps. This hybrid approach preserves the diversity benefits of the teacher model while still achieving significant computational savings.

Algorithm~\ref{alg:teacher_guided_sampling} outlines our teacher-guided sampling procedure. The process begins with random Gaussian noise $\mathbf{z}$ and progressively denoises it through a sequence of steps. The first $K$ steps are performed by the teacher model using a flow matching formulation with the sway sampling schedule \cite{chen2024f5}, which allocates more samples to early time steps where most of the semantic structure is established. Once the noise level reaches $t_{\text{switch}}$, the algorithm transitions to the student model, which completes the remaining denoising in just $M$ steps (typically 2-3). A key advantage of our approach is that it achieves a more favorable trade-off between computational efficiency and output diversity. By delegating the labor-intensive task of establishing prosodic structure to the teacher model and the refinement of acoustic details to the student model, we leverage the strengths of both approaches. The teacher model is employed for fewer steps than its typical full inference (approximately 6-14 steps instead of 32), while the student model still performs only a small number of denoising steps (2-3 instead of 4).

Our empirical evaluation (Table~\ref{tab:main_results}) confirms that teacher-guided sampling successfully mitigates the mode shrinkage problem, restoring the diversity of the generated speech to levels comparable to the teacher model, particularly in terms of pitch variation and cadence diversity. Notably, this improvement comes with only a modest increase in computational cost compared to the pure student model but still $1.8 \times$ faster than the full teacher model. Additionally, similar to the student model, our hybrid approach produces samples with better SIM and WER than the teacher-only samples, benefiting from the direct metric optimization of the DMOSpeech framework. 

The parameters $K$, $t_{\text{switch}}$, and $M$ offer flexible control over the trade-off between computational efficiency and output diversity. For applications where diversity is critical, such as creative content production, a higher $t_{\text{switch}}$ value (around 0.4-0.5) can be used, allocating more steps to the teacher model. Conversely, for applications where efficiency is paramount, such as real-time systems, a lower $t_{\text{switch}}$ value (around 0.1-0.2) can be employed with minimal degradation in perceptual quality.

\section{Experiments}
\begin{table*}[t]
\scriptsize
\setlength{\tabcolsep}{3pt}  % tighten inter-column spacing
\centering
\caption{Objective and subjective evaluation results on \textit{Seed-TTS-en} and \textit{Seed-TTS-zh} evaluation sets. 
CMOS-S and CMOS-N refer to CMOS for similarity and naturalness, respectively, with DMOSpeech\,2 (our system with 4 sampling steps) as the anchor (negative means DMOSpeech\,2 is better). The \textbf{best} values for objective evaluations are shown in \textbf{bold} and the \underline{second-best} values are \underline{underlined} where S/A stands for the same as above. For subjective evaluations, the statistically significant results are marked by one asterisk if $p < 0.05$ and two asterisks if $p < 0.01$. $\text{CV}_{f_0}$ is computed with the DDPM sampler for fairness.  }
\label{tab:main_results}
\resizebox{\linewidth}{!}{%
\begin{tabular}{l|cc|cc|cc|cc|c|c}
\toprule
\multirow{2}{*}{Model} 
& \multicolumn{2}{c|}{\makecell{\textit{Seed-TTS-en}}} 
& \multicolumn{2}{c|}{\makecell{\textit{Seed-TTS-zh}}} 
& \multicolumn{2}{c|}{\makecell{English}} 
& \multicolumn{2}{c|}{\makecell{Chinese}} 
& \multirow{2}{*}{$\text{CV}_{f_0}\uparrow$} 
& \multirow{2}{*}{RTF$\downarrow$} \\
& WER$\downarrow$ & SIM$\uparrow$ 
& CER$\downarrow$ & SIM$\uparrow$ 
& CMOS-N & CMOS-S
& CMOS-N & CMOS-S
& & \\
\midrule
Ground Truth 
& 2.143 & 0.734 
& 1.254 & 0.755 
& 0.03  & $-0.13^{*}$ 
& 0.02  & $-0.06$ 
& — & — \\
\midrule
F5-TTS Teacher (32 steps) 
& 1.947 & 0.662 
& 1.695 & 0.750 
& $-0.12^{*}$ & $-0.04$ 
& $-0.09$ & $-0.11^{*}$ 
& \textbf{0.6659} & 0.1671 \\
\hdashline
DMOSpeech\,2 (4 steps) 
& \underline{1.752} & \underline{0.698} 
& \underline{1.527} & \textbf{0.760} 
& 0.0 & 0.0 
& 0.0 & 0.0 
& 0.4640 & \textbf{0.0316} \\
\quad w/o duration predictor RL
& 3.750 & 0.672 
& 2.000 & 0.750 
& $-0.43^{**}$ & $-0.48^{**}$
& $-0.26^{*}$ & $-0.31^{*}$ 
& S/A & S/A \\
\hdashline
Teacher-Guided (16 steps) 
& \textbf{1.738} & \textbf{0.699} 
& \textbf{1.468} & \textbf{0.760} 
& 0.01 & $-0.03$ 
& $0.45^{**}$ & $0.3^{*}$ 
& \underline{0.5932} & \underline{0.0941} \\
\bottomrule
\end{tabular}}
\end{table*}

\subsection{Experimental Setup}

\textbf{Datasets}\quad Following F5-TTS \cite{chen2024f5}, we utilize the in-the-wild multilingual speech dataset Emilia \cite{He2024EmiliaAE} to train our models. After filtering out transcription failures and misclassified language speech, we retain approximately 95k hours of English and Chinese data. For evaluation, we adopt three test sets: Seed-TTS \cite{Anastassiou2024SeedTTSAF}  \textit{test-en} with 1088 samples from CommonVoice \cite{ardila2019common}, and Seed-TTS \textit{test-zh} with 2020 samples from DiDiSpeech\cite{guo2021didispeech}.

\noindent \textbf{Training}\quad For our teacher model, we adopt F5-TTS \cite{chen2024f5} with approximately 300M parameters, trained for 2M steps on the Emilia dataset. We maintain the same hyperparameter configuration as in the original F5-TTS, with a batch size of 307,200 audio frames (0.91 hours), using the AdamW optimizer \cite{loshchilov2018fixing} with a peak learning rate of 7.5e-5, linear warmup for 20K updates, and linear decay afterwards. For the student model training in DMOSpeech 2, we follow the approach in \cite{li2024dmospeech} but use half the batch size of the teacher model training. The learning rate for the student model resumes from the final learning rate of the teacher model training (around 6e-5) and continues for an additional 200K steps on the Emilia dataset. The duration predictor uses an encoder-decoder transformer architecture similar to DiTTo-TTS \cite{lee2024ditto}. It is initially trained on the Emilia dataset for 85K steps with a learning rate of 1e-4 and the same batch size as the F5-TTS teacher training. We use the AdamW optimizer with default parameters of Pytorch. After this initial training, we further fine-tune the duration predictor using GRPO \cite{sun2025f5r} for an additional 1.5K steps with a group size of 16, optimizing directly for speaker similarity and word error rate metrics. All experiments were conducted on 8 NVIDIA H100 GPUs.

\noindent \textbf{Baselines}\quad We compare several configurations of our models with both subjective and objective evaluations: (1) The ground truth recordings, (2) F5-TTS teacher without a duration predictor using 32 sampling steps, (3) DMOSpeech 2 with the RL-optimized duration predictor using 4 sampling steps, (4) student with the duration predictor before RL using 4 sampling steps, and (5) a teacher-guided sampling approach where the teacher model handles initial denoising steps before transitioning to the student model ($t_{switch} = 0.25$, with teacher handling 14 steps and student handling 2 steps, for a total of 16 steps). We use the pretrained Vocos vocoder \cite{siuzdak2023vocos} to convert generated mel-spectrograms to audio signals. We also compare our DMOSpeech 2 with several state-of-the-art TTS systems on objecetive metrics: CosyVoice 2 \cite{du2024cosyvoice2}, Spark-TTS \cite{wang2025spark}, LLaSA-8B \cite{ye2025llasa}, MaskGCT \cite{wang2024maskgct}, and our F5-TTS teacher model (32 steps) \cite{chen2024f5}. All samples were resampled to 24 kHz for a fair comparison. 

\subsection{Evaluation Metrics}

We evaluate our models under the \textit{cross-sentence} task, following the protocol established in \cite{le2024voicebox}. In this task, the model is given a reference text, a short speech prompt, and its transcription, and is required to synthesize speech reading the reference text while mimicking the voice characteristics of the prompt speaker.

For objective evaluation, we report word error rate (WER) and speaker similarity between generated and original target speeches (SIM). For WER, we employ Whisper-large-v3 \cite{radford2023robust} to transcribe English and Paraformer-zh \cite{gao2023funasr} for Chinese, following the approach in Seed-TTS \cite{Anastassiou2024SeedTTSAF}. For SIM-o, we use a WavLM-large-based \cite{chen2022wavlm} speaker verification model to extract speaker embeddings for calculating the cosine similarity between synthesized and ground truth speeches. We also measure the real-time factor (RTF) to evaluate inference speed, defined as the ratio of speech generation time to the duration of the generated speech on a single H100 GPU. Additionally, to demonstrate that teacher-guided sampling helps improve sampling diversity, we compare the coefficient of variation of the pitch ($\text{CV}_{f_0}$) of 50 different samples synthesized with the same input text and prompt across 20 text-prompt pairs for various configurations of our models averaged across all frames (with the same input total duration). For the teacher, we used DDPM \cite{ho2020denoising} modified for flow-matching \cite{gao2025diffusionmeetsflow} to have a fair comparison with the students as they have additional noise injections throughout the sampling process (see Algorithm \ref{alg:dmd_sampling} for more details). 

For subjective evaluation, we conduct human listening tests using comparative mean opinion scores (CMOS) for both naturalness and similarity. For CMOS, human evaluators are presented with randomly ordered synthesized speech from one model and an anchor model (our DMOSpeech 2 with the RL-optimized duration predictor using 4 sampling steps), and are asked to rate which sample has higher similarity with respect to the prompt speech and more like a human recording (either $+1$ or $-1$). We report the average scores of a total of 320 samples in both English and Chinese. For more details, we refer the readers to Appendix \ref{app:C}. 
\subsection{Results}

\subsubsection{\textbf{Main Results}}

Table \ref{tab:main_results} shows that DMOSpeech 2 with the RL-optimized duration predictor significantly outperforms both the teacher model and the student model without duration predictor optimization. On the English evaluation set, DMOSpeech 2 achieves a WER of $1.752$ compared to $1.947$ for F5-TTS and $3.750$ for DMOSpeech without RL optimization. For speaker similarity, DMOSpeech 2 reaches $0.698$ compared to $0.662$ for F5-TTS (teacher) and $0.672$ for DMOSpeech without RL. We observe similar improvements on the Chinese evaluation set, where DMOSpeech 2 achieves a CER of $1.527$ and similarity of $0.760$, outperforming both F5-TTS with $1.695$ CER and $0.750$ SIM, and DMOSpeech without RL with $2.000$ CER and $0.750$ SIM.

Most remarkably, DMOSpeech 2 delivers this superior performance while maintaining exceptional computational efficiency, with an RTF of $0.0316$, which is more than $5\times$ faster than the teacher model's $0.1671$. The teacher-guided sampling approach achieves slightly better objective metrics with WER of $1.738$ and CER of $1.468$ but an increased computation time.

The subjective CMOS evaluation further confirms our approach's effectiveness. Human evaluators rated DMOSpeech 2 significantly better than that without RL (i.e., the original DMOSpeech), with substantial margins in both English and Chinese. For English, DMOSpeech 2 showed naturalness superiority with CMOS-N of $-0.43$ and similarity advantage with CMOS-S of $-0.48$, both statistically significant at $p<0.01$. For Chinese, we observed similar benefits with CMOS-N of $-0.26$ and CMOS-S of $-0.31$, significant at $p<0.05$. DMOSpeech 2 also outperforms F5-TTS, achieving significantly better English naturalness with CMOS-N of $-0.12$ and Chinese similarity with CMOS-S of $-0.11$, both at $p<0.05$. Interestingly, while the teacher-guided sampling approach shows comparable performance to DMOSpeech 2 for English, it demonstrates significantly better subjective scores for Chinese, with CMOS-N reaching $+0.45$ at $p<0.01$ and CMOS-S of $+0.3$ at $p<0.05$. Perhaps most importantly, DMOSpeech 2 achieves results statistically indistinguishable from ground truth recordings in naturalness for both English and Chinese. For English similarity, it even achieves a noteworthy CMOS-S of $-0.13$ compared to ground truth, significant at $p<0.05$. These results confirm that our approach produces speech that approaches human-level quality on the evaluation benchmark dataset while maintaining exceptional computational efficiency.

\begin{figure}[!t]
  \centering
  \subfigure{\includegraphics[width=0.49\columnwidth]{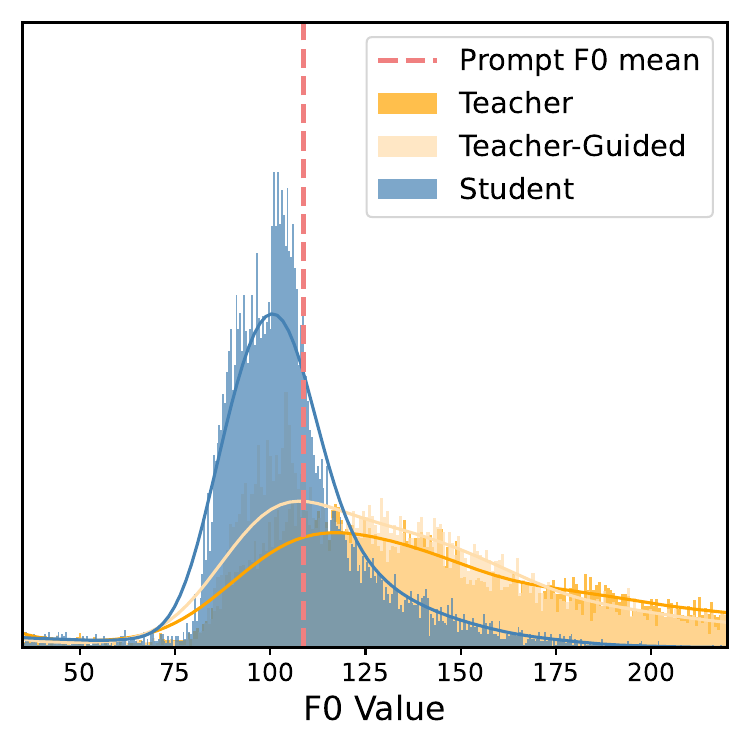}}
  \subfigure{\includegraphics[width=0.49\columnwidth]{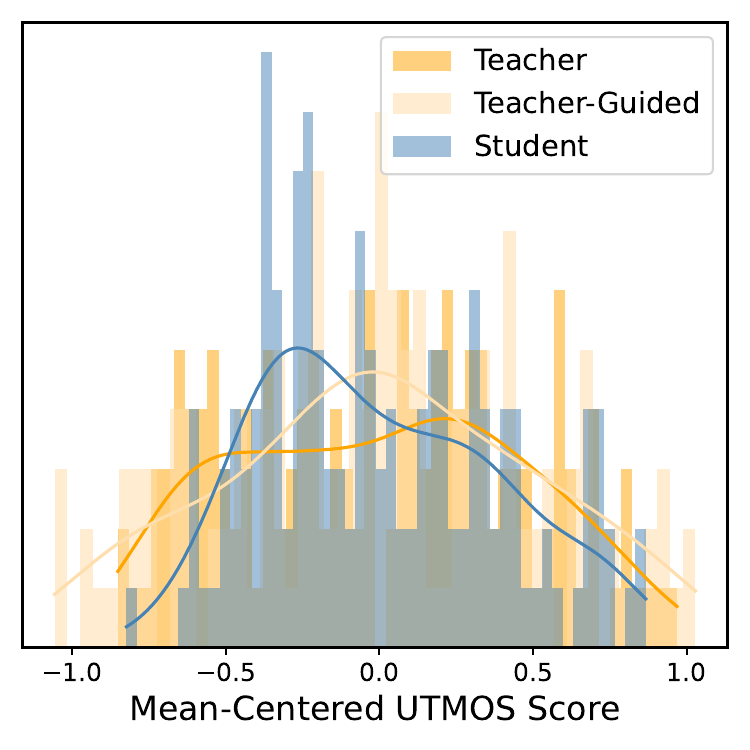}}
  \vspace{-10pt}
  \caption{Comparison of diversity across sampling methods. (a) F0 value distributions (shown as histograms and kernel density estimates). The student model (light blue) exhibits a much narrower distribution compared to the teacher model (yellow), indicating mode shrinkage in prosodic patterns. The teacher-guided approach (orange) successfully recovers much of this diversity. (b) Mean-centered UTMOS score distributions. Acoustic quality remains consistent across all models despite differences in prosodic diversity, supporting our hypothesis that diversity reduction primarily affects prosodic and temporal aspects rather than acoustics.}
  \label{fig:diversity_comparison}
\vspace{-10pt}
\end{figure}

\begin{table*}[!t]
\scriptsize
\centering

\caption{Comparison with state-of-the-art models on \textit{Seed-TTS-en} and \textit{Seed-TTS-zh} evaluation sets. The \textbf{best} values in each column are shown in \textbf{bold} and the \underline{second-best} values are \underline{underlined}. All samples from baseline models were synthesized using the official checkpoints released by the authors. }
\label{tab:sota_comparison}
\begin{tabular}{l|c|c|cc|cc|c}
\toprule
\multirow{2}{*}{Model} & \multirow{2}{*}{\#Params} & \multirow{2}{*}{Dataset (\# Hours)} & \multicolumn{2}{c|}{\textit{Seed-TTS-en}} & \multicolumn{2}{c|}{\textit{Seed-TTS-zh}} & \multirow{2}{*}{RTF$\downarrow$} \\
 & & & WER$\downarrow$ & SIM$\uparrow$ & CER$\downarrow$ & SIM$\uparrow$ & \\
\midrule
Ground Truth & -- & -- & 2.143 & 0.734 & 1.254 & 0.755 & -- \\
\midrule
F5-TTS (32 steps) \cite{chen2024f5} & 0.3B & Emilia \cite{He2024EmiliaAE} (95k hrs) & {1.947} & 0.662 & 1.695 & 0.750 & 0.167 \\
CosyVoice 2 \cite{du2024cosyvoice2} & 0.5B & Proprietary (200k hrs) & 3.358 & 0.641 & 1.582 & 0.754 & 0.527 \\
Spark-TTS \cite{wang2025spark} & 0.5B & VoxBox \cite{wang2025spark} (100k hrs) & 2.308 & 0.572 & {1.717} & 0.657 & 1.784 \\
MaskGCT \cite{wang2024maskgct} & 0.7B & Emilia \cite{He2024EmiliaAE} (95k hrs) & 2.622 & \textbf{0.713} & 2.395 & \textbf{0.772} & 2.397 \\
LLaSA-8B \cite{ye2025llasa} & 8B & Proprietary (200k hrs) & 3.994 & 0.594 & 4.214 & 0.671 & 1.374 \\
\midrule
DMOSpeech 2 (Student-Only, 4 steps) & 0.3B & Emilia \cite{He2024EmiliaAE} (95k hrs) & \underline{1.752} & {0.698} & \underline{1.527} & \underline{0.760} & \textbf{0.032} \\
DMOSpeech 2 (Teacher-Guided, 16 steps) & 0.6B & Emilia \cite{He2024EmiliaAE} (95k hrs) & \textbf{1.738} & \underline{0.699} & \textbf{1.468} & \underline{0.760} & \underline{0.094} \\
\bottomrule
\end{tabular}
\end{table*}

\subsubsection{\textbf{Comparison with State-of-the-Art Models}} Table \ref{tab:sota_comparison} show the comparison of DMOSpeech 2 with previous state-of-the-art TTS models on the Seed-TTS evaluation sets. DMOSpeech 2, in both its student-only and teacher-guided variants, significantly outperforms most baseline models in terms of intelligibility while maintaining competitive speaker similarity and vastly superior computational efficiency. Our student-only DMOSpeech 2 model achieves an English WER of $1.752$ and a Chinese CER of $1.527$, substantially better than all baseline models with similar or larger parameter counts. The next best performer, our teacher model F5-TTS, achieves a WER of $1.947$ and CER of $1.695$ with the same parameter count but requires $5.3\times$ more computation time. The teacher-guided variant further improves these results to $1.738$ WER and $1.468$ CER while still maintaining a $1.8\times$ speed advantage over the teacher model F5-TTS despite requiring twice the parameter size (from 0.3B to 0.6B), as it needs the weight of both the teacher and the student models. In terms of speaker similarity, DMOSpeech 2 variants score $0.698$-$0.699$ for English and $0.760$ for Chinese, outperforming most baselines except MaskGCT, which achieves the highest similarity scores but at the cost of significantly worse intelligibility and dramatically higher computational requirements. MaskGCT has an RTF of $2.397$, making it $75\times$ slower than DMOSpeech 2.

It is noteworthy that DMOSpeech 2 outperforms much larger models like LLaSA-8B  across all metrics, despite having only $0.3$B parameters compared to $8$B. This demonstrates that our targeted optimization approach through reinforcement learning of the duration predictor is more effective than simply scaling up model size. The computational efficiency of DMOSpeech 2 is particularly striking, with an RTF of $0.032$ for the student-only variant, making it $5.2\times$ faster than F5-TTS, $16.5\times$ faster than CosyVoice 2, $55.8\times$ faster than Spark-TTS, and $42.9\times$ faster than LLaSA-8B. This exceptional efficiency makes DMOSpeech 2 particularly suitable for real-time applications and deployment on resource-constrained devices.

\subsubsection{\textbf{Effect of Teacher-Guided Sampling on Diversity}}
As shown in Table~\ref{tab:main_results}, teacher-guided sampling successfully addresses diversity limitations in our distilled student model. The coefficient of variation of pitch ($\text{CV}_{f_0}$) reveals the teacher model's superior diversity (0.6659) compared to the student model's reduced variation (0.4640, a 30.3\% decrease), indicating the student model suffers from mode shrinkage. Our teacher-guided approach recovers much of this diversity (0.5932, 89.1\% of teacher's diversity) while maintaining superior WER and speaker similarity from the student model with direct metric optimization. Figure~\ref{fig:diversity_comparison}a illustrates this effect through F0 distributions. The student model shows a narrower, more peaked distribution than the teacher model, demonstrating mode shrinkage from aggressive step reduction. The teacher-guided approach successfully broadens this distribution. In Figure~\ref{fig:diversity_comparison}b, we plot the mean-centered UTMOS score distributions since different models demonstrate significant differences in their mean UTMOS scores. Despite this, the mean-centered distributions after remain consistent across all models, indicating diversity reduction occurs primarily in prosodic aspects rather than spectral characteristics. This hybrid approach achieves a favorable trade-off between computational efficiency (RTF = 0.0941) and output diversity by leveraging the teacher model for establishing prosodic structure and the student model for efficient acoustic refinement.

\section{Conclusion}

This paper introduces DMOSpeech 2, which addresses two critical limitations in end-to-end diffusion-based TTS systems: optimizing the duration predictor component for perceptual metrics and mitigating diversity reduction in distilled models. Through reinforcement learning with GRPO, we optimize the duration predictor directly for speaker similarity and intelligibility, while our teacher-guided sampling approach restores prosodic diversity. Comprehensive evaluations show that DMOSpeech 2 significantly outperforms previous state-of-the-art models across various metrics while maintaining exceptional computational efficiency. The ability to optimize the previously isolated duration predictor component marks significant progress in end-to-end TTS optimization. Future work could explore applying our targeted RL approach to other components in generative pipelines that are difficult to optimize directly with gradient descent, such as the teacher model in the hybrid sampling approach, and employing rewards other than WER and SIM to align our models with human perceptions further.  

DMOSpeech 2 raises important societal considerations. Our system's improved speaker similarity and intelligibility offer significant benefits for accessibility, personalized assistants, and content creation. However, like all high-fidelity voice synthesis technologies, it presents potential risks for voice spoofing and deepfakes. The computational efficiency of our approach also democratizes access to this technology, amplifying both benefits and risks. To address these concerns, we emphasize the importance of developing more robust detection methods for synthetic speech and establishing appropriate governance frameworks. To foster further research and reproducibility, we will release our source code and pre-trained models publicly. We believe our open-source approach will accelerate progress in addressing both the technical challenges and ethical considerations associated with advanced TTS systems.

\bibliography{mybib}
\bibliographystyle{unsrt}

\newpage

%%%%%%%%%%%%%%%%%%%%%%%%%%%%%%%%%%%%%%%%%%%%%%%%%%%%%%%%%%%%

\appendix

\section{Additional Analyses}
\label{app:A}
\subsection{Impact of Duration Prediction on Speech Quality}
\vspace{-10pt}
\begin{table}[!ht]
\centering
\caption{Impact of different duration prediction approaches on speech quality metrics. All evaluations are conducted on \textit{Seed-TTS-en} dataset using the same speech generation model.}
\label{tab:duration_ablation}
\begin{tabular}{l|c|c}
\toprule
Duration Source & SIM$\uparrow$ & WER$\downarrow$ \\
\midrule
Ground Truth Audio & 0.734 & 2.143 \\
\midrule
Ground Truth Duration & 0.697 & 1.821 \\
Speaking Rate Based & 0.682 & 2.028 \\
Duration Predictor & 0.672 & 3.750 \\
Best-of-8 Sampling & 0.724 & 1.723 \\
\midrule
DMOSpeech 2 (Ours) & 0.698 & 1.752 \\
\bottomrule
\end{tabular}
\end{table}

Duration prediction plays a crucial role in non-autoregressive TTS systems, directly affecting both intelligibility and speaker similarity. To illustrate this impact, we conducted experiments comparing different duration determination methods on the \textit{Seed-TTS-en} evaluation set. Table \ref{tab:duration_ablation} presents the results.

We evaluated several approaches to determine speech duration: Ground Truth Audio refers to the original recordings; Ground Truth Duration uses reference durations from the dataset; Speaking Rate Based implements the F5-TTS approach of interpolating duration based on speaking rate; Duration Predictor shows results without RL optimization; Best-of-8 Sampling selects the best result from 8 different duration samples based on quality metrics; and DMOSpeech 2 features our proposed RL-optimized duration predictor.

The results demonstrate several important findings. First, the unoptimized duration predictor performs notably worse than other approaches, particularly in terms of intelligibility (WER of 3.750). This confirms our hypothesis that duration prediction is a critical bottleneck in TTS quality, which has also been shown in previous studies \cite{eskimez2024e2}. 

Second, the Best-of-8 sampling approach achieves the best results with a WER of 1.723 and SIM of 0.724. This represents an "oracle" upper bound on what could be achieved through effective duration prediction, as it leverages privileged information about outcome quality that would not be available during standard inference. This ceiling indicates the theoretical limit of what our RL approach could achieve with perfect optimization.

Notably, our proposed RL-optimized duration predictor (DMOSpeech 2) achieves a WER of 1.752, which is better than using ground truth durations (WER of 1.821) while maintaining competitive similarity (0.698). This demonstrates that our RL-based optimization successfully learns to predict durations that enhance speech intelligibility without requiring ground truth information. While not quite reaching the ceiling established by Best-of-8 sampling, our approach comes remarkably close while being significantly more efficient, requiring only a single forward pass during inference.

Interestingly, while using ground truth durations provides good intelligibility, it does not maximize speaker similarity (SIM of 0.697). This suggests that optimal durations for speaker similarity might differ slightly from those for intelligibility, highlighting the benefit of our joint optimization approach through reinforcement learning, which can balance these competing objectives.

\subsection{Hyperparameter Selection for Duration Predictor RL}
\label{app:rl_hyperparams}

\subsubsection{\textbf{Group Size and Training Steps}} 
\begin{figure*}[!ht]
\centering
\includegraphics[width=0.95\textwidth]{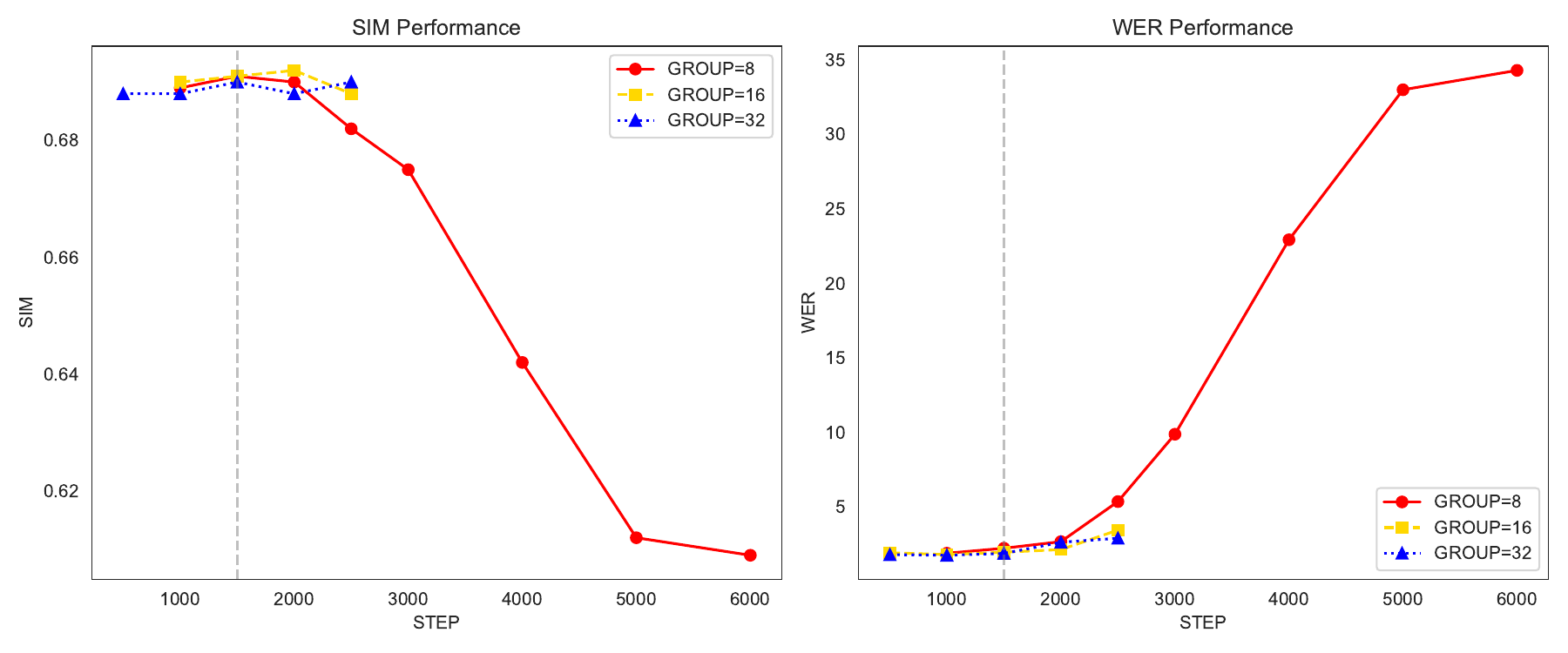}
\caption{Performance dynamics during RL training of the duration predictor with various group sizes. The left plot shows speaker similarity (SIM) while the right plot shows word error rate (WER). The vertical dashed line indicates the 1.5K steps we selected for our final model.}
\label{fig:rl_training_dynamics}
\end{figure*}

To determine the optimal hyperparameters for our GRPO-based duration predictor training, we conducted extensive validation experiments using a small subset of the \textit{Seed-TTS-en} evaluation set. Figure~\ref{fig:rl_training_dynamics} illustrates the dynamics of model performance across different training steps and group sizes.

Our experiments revealed a critical training steps threshold around 1.5K steps, beyond which performance deteriorated significantly. With a group size of 8, both the speaker similarity and word error rate metrics showed dramatic degradation after approximately 2K steps. This pattern suggests that extended training with reinforcement learning leads to overfitting to the reward signal, causing the policy to deviate excessively from the reference model.

Interestingly, while larger group sizes (16 and 32) demonstrated greater stability in performance over extended training, group size 16 emerged as the optimal configuration. We hypothesize that this superiority stems from group size 16 achieving an ideal balance between exploration and exploitation. With 16 samples per training instance, the model receives sufficient diversity in speech realizations to explore the duration space effectively, while maintaining enough focus on high-reward regions to exploit promising speech characteristics. 

Additionally, group size 16 provides adequate statistical stability for reliable advantage estimation without introducing excessive computational overhead. Smaller groups (8) appear to suffer from high variance in advantage estimation, leading to unstable training, while larger groups (32) offer diminishing returns in performance improvement relative to their increased computational cost.

Based on these findings, we selected 1.5K training steps with a group size of 16 for our final model, which strikes an optimal balance between performance improvement and training efficiency. This configuration effectively improves the duration predictor's accuracy without deviating too far from the original supervised model, thereby avoiding the pitfalls of reward over-optimization.

\subsubsection{\textbf{Balancing Speaker Verification and Speech Recognition Rewards}}
\label{app:reward_balancing}

A critical aspect of our reinforcement learning approach is properly balancing the contributions of speaker similarity and speech intelligibility in the reward function. Our reward formulation combines a speaker verification (SV) similarity term and a connectionist temporal classification (CTC) likelihood term:

\begin{equation}
    r_k = \log p(\mathbf{x} | C(\mathbf{y}_k)) + \lambda_{\text{SIM}} \cdot \frac{\mathbf{e}_{\text{p}} \cdot \mathbf{e}_{y_k}}{\norm{\mathbf{e}_{\text{p}}} \norm{\mathbf{e}_{y_k}}},
\end{equation}

The selection of an appropriate $\lambda_{\text{SIM}}$ value is crucial for ensuring that neither component dominates the optimization process. During our preliminary analysis, we observed that the CTC term ($\log p(\mathbf{x} | C(\mathbf{y}_k))$) typically produces values approximately three times larger in magnitude than the cosine similarity term with our CTC and SV models. This imbalance would naturally lead the duration predictor to prioritize intelligibility over speaker mimicry if left unaddressed.

To achieve a balanced optimization objective where both metrics contribute equally to model training, we conducted a series of calibration experiments. By analyzing the statistical distribution of both reward components across our validation set, we determined that setting $\lambda_{\text{SIM}} = 3$ effectively equalizes their contributions. This calibration ensures that improvements in speaker similarity receive comparable reinforcement to improvements in speech intelligibility.

Our experimental results confirm the effectiveness of this balanced approach. When using significantly lower values for $\lambda_{\text{SIM}}$, we observed that the model would converge to durations that produced more intelligible speech but with diminished speaker similarity. Conversely, with substantially higher values, the model prioritized speaker characteristics at the expense of comprehensibility. The selected value of $\lambda_{\text{SIM}} = 3$ achieves the optimal trade-off between these competing objectives, resulting in speech that maintains both high intelligibility and strong speaker similarity.

\section{DMOSpeech Technical Details}
\label{app:B}
This section provides a comprehensive overview of the DMOSpeech framework \citep{li2024dmospeech} as adapted for flow matching models in DMOSpeech 2.

\subsection{Flow Matching for Speech Synthesis}
\label{appA.1}

Our teacher model, F5-TTS \citep{chen2024f5}, is based on the conditional flow matching (CFM) framework rather than the velocity prediction diffusion used in the original DMOSpeech. The flow matching objective is to match a probability path $p_t$ from a simple distribution $p_0$ (standard normal) to a target distribution $p_1$ that approximates the data distribution $q$.

In the CFM framework, the model learns a vector field $v_t$ that guides the transformation of samples from noise to data. The loss function is:
\begin{equation}
\mathcal{L}_{\text{CFM}}(\theta) = \mathbb{E}_{t, q(x_{1}), p(x_0)} \| v_{t}((1-t)x_0+tx_1) - (x_1-x_0) \| ^2
\end{equation}
\noindent where $t \sim \mathcal{U}[0,1]$ is the flow step, $x_0 \sim p(x_0)$ is sampled from the noise distribution, $x_1 \sim q(x_1)$ is sampled from the data distribution, and $(1-t)x_0+tx_1$ represents the noisy sample at time $t$.

For speech synthesis, the input consists of a mel spectrogram $x_1 \in \mathbb{R}^{F \times N}$ where $F$ is the mel dimension and $N$ is the sequence length; a text embedding $\mathbf{c}$ derived from the input text; and a binary mask $\mathbf{m} \in \{0,1\}^{F \times N}$ that indicates which portions are prompt (to be preserved) and which are to be generated

The model $v_t$ is trained to predict the flow vector field conditioned on these inputs. During training, we introduce a noisy sample $(1-t)x_0 + tx_1$ and the masked speech $(1-\mathbf{m}) \odot x_1$, where $x_0$ is Gaussian noise.

During inference, we use an ordinary differential equation (ODE) solver to transform noise $x_0$ into a mel spectrogram $x_1$ by integrating along the vector field:
\begin{equation}
\frac{d\psi_t(x_0)}{dt} = v_t(\psi_t(x_0)|\mathbf{c}, \mathbf{m})
\end{equation}
\noindent where $\psi_0(x_0) = x_0$ and we aim to compute $\psi_1(x_0) = x_1$.

\subsection{Sway Sampling for Improved Inference}
\label{appA.2}

F5-TTS \citep{chen2024f5} introduced sway sampling to improve the efficiency and quality of speech generation. The sway sampling function is defined as:
\begin{equation}
f_{\text{sway}}(u; s) = u + s \cdot (\cos(\frac{\pi}{2}u) - 1 + u)
\end{equation}
\noindent where $u \sim \mathcal{U}[0,1]$ and $s$ is a coefficient controlling the sampling bias. This function transforms uniform samples to focus more on certain flow regions.

In DMOSpeech 2, we use a specific sway sampling schedule with the coefficient $s=-1$ that transforms our standard 4-step schedule $\{0.0, 0.25, 0.5, 0.75\}$ to $\{0.0000, 0.0761, 0.2929, 0.6173\}$ following \cite{chen2024f5}. This places more emphasis on the early steps of the generation process, allowing the model to establish better content and speaker foundations before refining details.

\subsection{Distribution Matching Distillation}
\label{appA.3}

DMOSpeech 2 adapts the Improved Distribution Matching Distillation (DMD 2) framework \citep{yin2024improved} for flow matching models. The objective is to train a student generator $G_\theta$ to produce samples whose distribution matches the data distribution after applying the forward flow process.

We minimize the Kullback-Liebler (KL) divergence between the distributions of the sampled real data $p_{\text{data}, t}$ and the sampled student generator outputs $p_{\theta, t}$ across all time $t \in [0,1]$:
\begin{align}
D_{KL}(p_{\theta, t} || p_{\text{data}, t} ) &= \mathbb{E}_{\mathbf{x} \sim p_{\theta, t} }\left[\log\left(\frac{p_{\theta, t}(\mathbf{x} )}{p_{\text{data}, t}(\mathbf{x} )}\right)\right] \nonumber \\
&= - \mathbb{E}_{\mathbf{x} \sim p_{\theta, t} }\left[ \log\left({p_{\text{data}, t}(\mathbf{x} )}\right) -  \log\left({p_{\theta, t}(\mathbf{x} )}\right)\right]
\end{align}
The DMD loss is defined as:
\begin{equation}
\mathcal{L}_{\text{DMD}} = {\mathbb{E}}_{t \sim \mathcal{U}(0, 1)} \left[D_{KL}(p_{\theta, t} || p_{\text{data}, t} ) \right]
\end{equation}
For flow matching models, we adapt the gradient formulation \cite{liu2024autoregressive}:
\begin{equation}
\nabla_\theta \mathcal{L}_{\text{DMD}} = 
-\mathop{\mathbb{E}}\limits_{\substack{t, \mathbf{x}_t, \mathbf{z}}} 
\left[
        \omega_t \left( v_{\text{real}}(\mathbf{x}_t, t) - v_{\theta}(\mathbf{x}_t, t) \right)\frac{dG}{d\theta} 
\right]
\end{equation}
\noindent where $\mathbf{x}_t = (1-t)G_\theta(\boldsymbol{z}) + t\mathbf{x}_1$ for $\mathbf{z} \sim \mathcal{N}(\mathbf{0}, \mathbf{I})$, and $v_{\text{real}}$ and $v_{\theta}$ are the vector fields from the teacher and student models, respectively. The weighting factor $\omega_t$ is defined as:
\begin{equation}
\omega_t = (1-t)
\end{equation}
\noindent which gives more weight to earlier flow steps, aligning with the sway sampling philosophy.

\subsection{Multi-step Sampling for Student Models}
\label{appA.4}

To address artifacts resulting from the one-step student model, we adapt the multi-step sampling approach from DMD 2 to the flow-matching model. The student generator $G_\theta$ is conditioned on the flow step $t$ to estimate the mel spectrogram from a noisy counterpart at predefined time steps $t \in \{t_1, \ldots, t_N\}$.

The multi-step sampling algorithm follows:

\begin{algorithm}[H]
\caption{DMD Multi-Step Sampling with Flow Matching}
\label{alg:dmd_sampling}
\begin{algorithmic}[1]
\Require Generator $G_\theta$, flow steps $\{t_1, \ldots, t_N\}$, text embedding $\mathbf{c}$, prompt mask $\mathbf{m}$
\State Sample $\mathbf{z} \sim \mathcal{N}(0, \mathbf{I})$
\State $\mathbf{x}_{t_1} \leftarrow \mathbf{z}$
\For{$n = 1$ to $N-1$}
    \State $\hat{\mathbf{x}}_1^n \leftarrow G_\theta(\mathbf{x}_{t_n}\,; \mathbf{c}, \mathbf{m}, t_n)$
    \State Sample $\boldsymbol{\epsilon} \sim \mathcal{N}(0, \mathbf{I})$
    \State $\mathbf{x}_{t_{n+1}} \leftarrow (1-t_{n+1}) \mathbf{z} + t_{n+1} \hat{\mathbf{x}}_1^n$
\EndFor
\State $\hat{\mathbf{x}}_1^N \leftarrow G_\theta(\mathbf{x}_{t_N}\,; \mathbf{c}, \mathbf{m}, t_N)$\\

\Return $\hat{\mathbf{x}}_1^N$
\end{algorithmic}
\end{algorithm}

This creates a progressive refinement process, where earlier steps establish the content and speaker characteristics while later steps add details.

\subsection{Multimodal Adversarial Training}
\label{appA.5}

To further improve the student model's performance, we incorporate adversarial training following \citet{yin2024improved}. Our discriminator $D$ is a conformer that takes as input the stacked features from all transformer layers of the student network with noisy input, along with the text embeddings $\mathbf{c}$, prompt mask $\mathbf{m}$, and flow step $t$ (denoted collectively as $\mathcal{C}$), adapted from \cite{li2024styletts}.

The adversarial loss functions are:

\begin{equation}
      \mathcal{L}_\text{adv} (G_\theta; D) = 
      \mathbb{E}_{\substack{t, \hat{\mathbf{x}}_t \sim p_{\theta, t}, \mathbf{m}}} \left[\left(D\left(\hat{\mathbf{x}}_t\,; \mathcal{C}\right) - 1\right)^2\right]
\end{equation}

\begin{align}
      \mathcal{L}_\text{adv}  (D; G_\theta) &= 
      \mathbb{E}_{t}\left[ \mathbb{E}_{\hat{\mathbf{x}}_t \sim p_{\theta, t }, \mathbf{m}}\left[\left(D\left(\hat{\mathbf{x}}_t\,; \mathcal{C}\right) \right)^2\right]\right] + \nonumber \\
      & \mathbb{E}_{t}\left[ \mathbb{E}_{{\mathbf{x}}_t \sim p_{\text{data}, t }, \mathbf{m}}\left[\left(D\left({\mathbf{x}}_t\,; \mathcal{C}\right) - 1\right)^2\right]\right]
\end{align}
\noindent where $\mathcal{C} = \{\mathbf{c}, \mathbf{m}, t\}$ and $\hat{\mathbf{x}}_t = (1-t) \mathbf{z} + t G_\theta(\mathbf{z}; \mathcal{C})$ for $\mathbf{z} \sim \mathcal{N}(0, I)$.

\subsection{Direct Metric Optimization}
\label{appA.6}

DMOSpeech 2 retains the direct metric optimization approach from the original DMOSpeech, allowing end-to-end optimization of perceptual metrics. We directly optimize both speaker embedding cosine similarity (SIM) and word error rate (WER).

For WER improvement, we incorporate a connectionist temporal classification (CTC) loss:
\begin{equation} 
\mathcal{L}_{\text{CTC}} = \mathbb{E}_{\mathbf{x}_{\text{fake}} \sim p_{\theta}, \mathbf{c}} \left[-\log p(\mathbf{c} | C(\mathbf{x}_{\text{fake}}))\right]
\end{equation}
\noindent where $\mathbf{x}_{\text{fake}}$ is the student-generated mel spectrogram, $\mathbf{c}$ is the text transcript, and $C(\cdot)$ is a pre-trained CTC-based ASR model operating on mel-spectrograms.

For speaker similarity, we use a speaker verification (SV) loss:
\begin{equation} 
\mathcal{L}_{\text{SV}} = \mathbb{E}_{\substack{ \mathbf{x}_{\text{real}} \sim p_{\text{data}}, \\ \mathbf{x}_{\text{fake}} \sim p_{\theta}, \mathbf{m}}}\left[1 - \frac{\mathbf{e}_{\text{real}} \cdot \mathbf{e}_{\text{fake}}}{\norm{\mathbf{e}_{\text{real}}} \norm{\mathbf{e}_{\text{fake}}}}\right]
\end{equation}
\noindent where $\mathbf{e}_{\text{real}} = {S}(\mathbf{x}_{\text{real}})$ and $\mathbf{e}_{\text{fake}} = {S}(\mathbf{x}_{\text{fake}})$ are the speaker embeddings of the prompt and student-generated speech, obtained from a pre-trained speaker verification model $S$.

\subsection{Training Objectives and Stability}
\label{appA.7}

The overall training objective for $G_\theta$ combines DMD, adversarial, SV, and CTC losses:

\begin{equation} 
\min_{\theta} \text{    } \mathcal{L}_{\text{DMD}} + \lambda_{\text{adv}} \mathcal{L}_{\text{adv}}(G_\theta;D) + \lambda_{\text{SV}} \mathcal{L}_{\text{SV}} + \lambda_{\text{CTC}} \mathcal{L}_{\text{CTC}}
\end{equation}

The training objectives for the student vector field model $g_{\boldsymbol{\psi}}$ and discriminator $D$ are:

\begin{equation}
\min_{\boldsymbol{\psi}} \text{    } \mathcal{L}_{\text{CFM}}\left(g_{\boldsymbol{\psi}}; p_{\theta}\right)
\end{equation}

\begin{equation}
\min_{D} \text{    } \mathcal{L}_{\text{adv}}\left(D; G_\theta\right)
\end{equation}

We employ an alternating training strategy where $G_\theta$, $g_{\boldsymbol{\psi}}$, and $D$ are updated at different rates to maintain stability. For every update of $G_\theta$, five updates of $g_{\boldsymbol{\psi}}$ are performed to ensure the vector field model can adapt quickly to changes in the generator distribution. The discriminator $D$ and generator $G_\theta$ are updated at the same rate.

For training stability, following \cite{li2024dmospeech}, the weights are set as follows: $\lambda_{\text{adv}} = 10^{-3}$ to balance the gradient norms, $\lambda_{\text{CTC}} = 0$ for the first 5,000 iterations, then $\lambda_{\text{CTC}} = 1$, and $\lambda_{\text{SV}} = 0$ for the first 10,000 iterations, then $\lambda_{\text{SV}} = 1$. This phased approach allows the generator to first learn basic speech generation before focusing on specific quality metrics.

\subsection{Vocoder}
\label{app:vocoder}

Same as F5-TTS \cite{chen2024f5}, DMOSpeech 2 uses the Vocos neural vocoder \citep{siuzdak2023vocos} to convert mel-spectrograms to waveforms. Vocos is a GAN-based vocoder that offers high-quality synthesis with efficient inference. The vocoder is pre-trained on a diverse dataset of speech recordings and is used as-is without fine-tuning during DMOSpeech 2 training and inference.

\subsection{Automatic Speech Recognition (ASR) Model}
\label{app:ASR}

The ASR model used for the CTC loss is a 6-layer transformer encoder trained directly on mel-spectrograms. The model is trained on Emilia using the CTC loss to align the speech with the text transcriptions for both Chinese and English.

\subsection{Speaker Verification (SV) Model}
\label{app:SV}

The speaker verification model is a 6-layer transformer encoder with an additional projection layer that produces fixed-dimensional speaker embeddings. The model is distilled from the WeSpeaker \cite{wang2023wespeaker} SimAMResNet34 model on the Emilia dataset following \cite{li2024dmospeech}.

\section{Subjective Evaluation}
\label{app:C}
\label{app:comparative_eval}

\begin{figure*}[!th]
    \centering
    \includegraphics[width=0.95\textwidth]{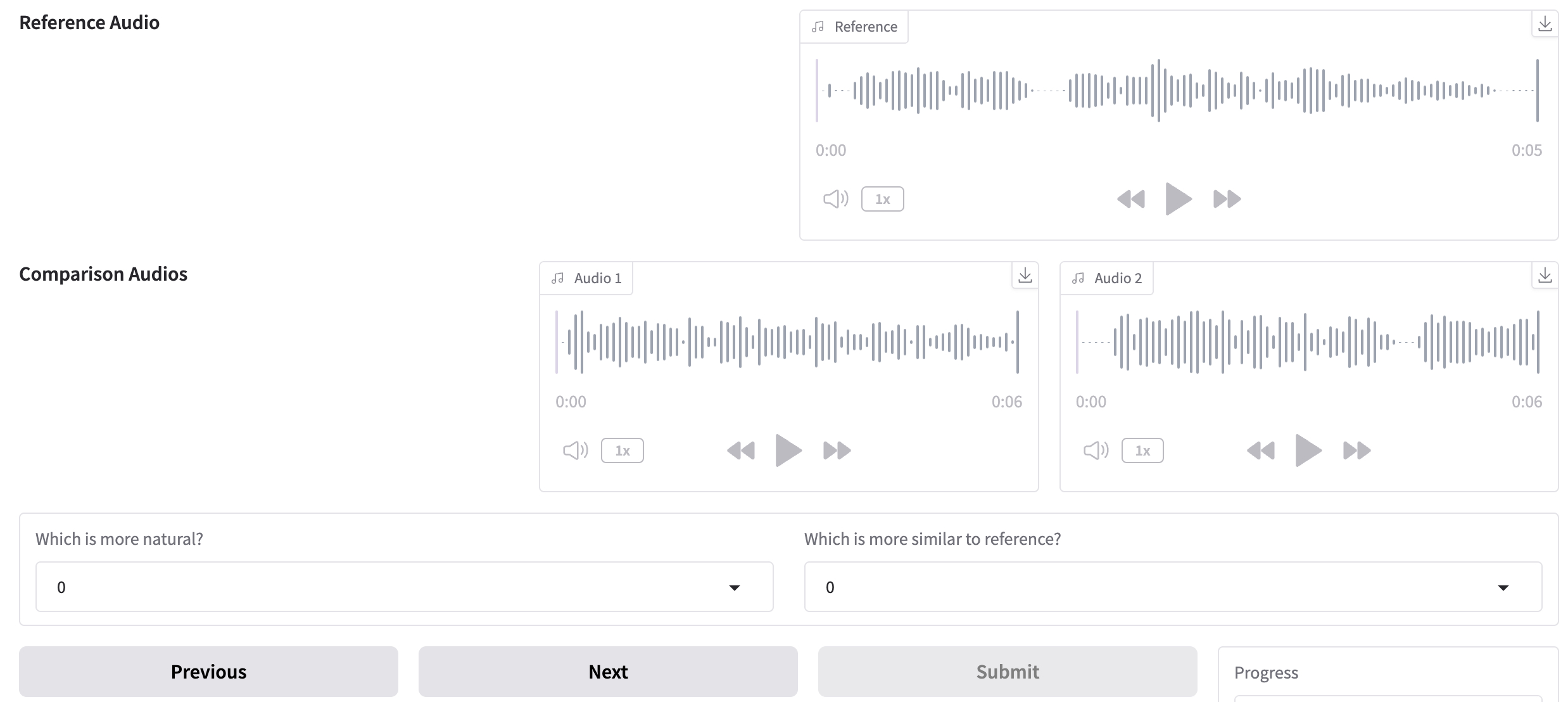}
    \caption{Screenshot of the comparative subjective evaluation interface. The interface presents three audio samples: a reference recording at the top, and two synthesized speech samples for comparison below. Participants are asked to make direct comparisons between the two synthesized samples by selecting which one sounds more natural and which one is more similar to the reference voice.}
    \label{fig:comparative_eval}
\end{figure*}

In addition to the absolute rating evaluation described previously, we conducted comparative mean opinion score (CMOS) tests to directly assess the relative performance of our proposed models against baseline systems. As shown in Figure~\ref{fig:comparative_eval}, the evaluation interface presents participants with three audio samples: a reference recording (top) and two synthesized speech samples (bottom) labeled as "Audio 1" and "Audio 2."

For each comparison, participants were instructed to:
\begin{enumerate}
    \item Listen to all three audio samples
    \item Select which of the two synthesized samples sounds more natural (left question)
    \item Select which of the two synthesized samples sounds more similar to the reference voice (right question)
\end{enumerate}

The DMOSpeech 2 model (with 4 sampling steps) served as the anchor system for all comparisons, with participants unaware of which sample corresponded to which system. The dropdown selection options were coded as follows: a rating of 0 indicates no preference, positive values indicate a preference for Audio 2, and negative values indicate a preference for Audio 1. This design allows for direct assessment of relative differences between systems without requiring absolute judgments on a fixed scale.

We collected responses from a total of 320 English and 320 Chinese samples. To ensure data quality, we employed validation checks similar to those in our previous evaluation, including mismatched speaker checks and identical sample pairs. Participants failing these validation tests were excluded from the final analysis. Statistical significance was determined using a paired t-test.

\end{document}